\documentclass[a4paper,11pt,dvips]{article}
\usepackage{graphicx}
\usepackage{amsmath}
\usepackage{amssymb}
\usepackage{latexsym}
\usepackage{color}
\newcommand{\bra}{\begin{array}}
\newcommand{\era}{\end{array}}
\newcommand{\beq}{\begin{equation}}
\newcommand{\eeq}{\end{equation}}
\newcommand{\bqr}{\begin{eqnarray}}
\newcommand{\eqr}{\end{eqnarray}}

\def\BC{\bb C}
\def\_\BC{\bbi C}

\def\( {\left(}
   \def\) {\right)}
\def\[ {\left[}
\def\] {\right]}
\def\no2 {{\textstyle{n\over 2}}}

\def\dag {{\dagger}}
\newcommand{\si}{\sigma}

\newcommand{\te}{\theta}

\begin{document}
\begin{titlepage}
\setcounter{page}{1}
\renewcommand{\thefootnote}{\fnsymbol{footnote}}

\begin{flushright}
%ucd-tpg 10-01\\
%arXiv:yymm.xxxx
\end{flushright}

\vspace{5mm}
\begin{center}

{\Large \bf {Confined Dirac Particles in Constant and Tilted Magnetic Field}}

\vspace{5mm}

{\bf Abdulaziz D. Alhaidari}$^{a}$,
{\bf Hocine Bahlouli}$^{a,b}$
and
{\bf Ahmed Jellal}$^{a,c,d}$\footnote{\sf ajellal@ictp.it -- a.jellal@ucd.ac.ma}

\vspace{5mm}

{$^a$\em Saudi Center for Theoretical Physics, Dhahran, Saudi Arabia}

{$^b$\em Physics Department,  King Fahd University
of Petroleum $\&$ Minerals,\\
Dhahran 31261, Saudi Arabia}

{$^c$\em Physics Department, College of Sciences, King Faisal University,\\
PO Box 380, Alahsa 31982, Saudi Arabia}

{$^{d}$\em Theoretical Physics Group,  %Department of Physics,
Faculty of Sciences, Choua\"ib Doukkali University},\\
{\em PO Box 20, 24000 El Jadida,
Morocco}

\vspace{3cm}

\begin{abstract}

We study the confinement of charged Dirac particles in 3+1 space-time due to the presence of a constant and tilted magnetic field.
We focus on the nature of the solutions of the Dirac equation and on how they depend on the choice of vector potential that gives
rise to the magnetic field. In particular, we select a "Landau gauge" such that the momentum is conserved along the direction of
the vector potential yielding spinor wavefunctions, which are localized in the plane containing the magnetic field and normal to
the vector potential. These wave functions are expressed in terms of the Hermite polynomials. We point out the relevance of these
findings to the relativistic quantum Hall effect and compare with the results obtained for a constant magnetic field normal to
the plane in 2+1 dimensions.
\vspace{3cm}



\end{abstract}
\end{center}
\end{titlepage}

%%%%%%%%%%%%%%%%%%%%%%%%%%%%%%%%%%%%%%%%%%%%%%
\section{Introduction}
%%%%%%%%%%%%%%%%%%%%%%%%%%%%%%%%%%%%%%%%%%%%%%%

 It is well known that the application of a magnetic field changes drastically the physical properties of the system and its energy spectra. 
 In particular, in a strong uniform magnetic field it results in a sequence of quantized Landau energy levels and associated wavefunctions 
 characterizing the system dynamics in the two-dimensional (2D) plane normal to the applied magnetic field. This quantization gives rise to 
 important phenomena in condensed matter physics, affects drastically the density of states and gives rise to the famous quantum Hall effect 
 (QHE) \cite{1}. The relativistic extension of these models turned out to be of great importance for the description of 2D quantum phenomena such
 as QHE in graphene \cite{2,22,222,2222}. Graphene is a single atomic sheet of graphite whose carriers have a linear dispersion relation at low energies
 \cite{3,33}. The recent surge of interest in this system originates from the analogy of its dispersion relation to that of relativistic, massless fermions, 
 hence leading to intriguing new phenomena \cite{4}. However, a layered system of graphene will generate a mass term for the carriers due to the interlayer
 coupling and hence will bring the problem to a massive Dirac equation. Due to these findings the study of the relativistic Dirac equation for massive particles 
 can be used to describe many of the low energy physical properties of graphene.

 In this paper, we investigate the Dirac equation in 3+1 dimensions and study the dynamics of a relativistic particle in a tilted constant magnetic field. 
 The details of our theoretical calculations and their simplicity depend heavily on the choice of the background gauge field giving rise to the magnetic field. 
 However, the gauge dependence of the various intermediate quantities calculated in this article should not distract us from the real requirement that measurable 
 physical quantities in the presence of a magnetic field are gauge invariant. For completeness, we perform in the appendix a similar computation in a different 
 gauge just to show how complicated the mathematics can be if we choose -so to speak- the wrong gauge. However, it is easy to establish the necessary unitary 
 transformation that enables one to go from the old gauge to the new one.

 %To make contact with graphene, 
 We study first the energy spectrum and associated eigenfunctions of the Dirac particles in a tilted magnetic field in 3+1 dimensions. 
 To achieve this goal we start by setting up our problem and performing different transformations to bring the solution into its final stage. By assuming separation 
 of variables we end up with a second order differential equation, which can be solved by using the well-known machinery of the quantum harmonic oscillator. The final 
 solutions are interesting because of their special dependence on the different physical parameters offering us the possibility to make a systematic study of various 
 graphene properties in the future. More specifically, the obtained energy spectrum is appealing because it gives an interesting dependence on the tilt angle of the 
 magnetic field and on the characteristic size of confinement. This will also help us shed some light on QHE in graphene and related matters where the angular 
 dependence will play a crucial role in this respect.

 The present paper is organized as follows. In section 2, we set up the problem by giving the necessary tools, which helps in obtaining a solvable problem. 
 In section 3, we introduce a coordinate transformation to ensure that the Jacobian is constant and simplify further our calculation. The obtained solutions 
 are summarized in a flowchart type of diagram, which shows the different steps used in order to obtain the eigenspinors. We discuss the nature of the energy
 spectrum obtained and emphasize two limits that are relevant to Dirac particles in graphene. Finally, we close by making some concluding remarks.

%%%%%%%%%%%%%%%%%%%%%%%%%%%%%%%%%%%%%%%%%%%%%
\section{Problem setting}
%%%%%%%%%%%%%%%%%%%%%%%%%%%%%%%%%%%%%%%%%%%%%%%%%
 To formulate the problem, we start by introducing the 3+1 dimensional Dirac equation in the unit
 system  $\hbar =c=1$, which then reads as 
\begin{equation} \label{GrindEQ__1_}
\gamma ^{\mu } (i\partial _{\mu } +A_{\mu } )\psi =m\psi .
\end{equation}
We choose the constant magnetic field in the \textit{x}-\textit{z} plane as $\vec{B}=(B_{x} ,0,B_{z} )=$ $B(\sin \theta ,0,\cos \theta )$. 
Moreover, the time component of the electrostatic potential is $A_{0} (\vec{r})=$ $V(x,z)$ while the space component in our chosen "Landau gauge" 
is written as
\begin{equation} \label{GrindEQ__2_}
\vec{A}(\vec{r})=(0,ax-bz,0)
\end{equation}
where $a=B_{z} =B\cos \theta $ and $b=B_{x} =B\sin \theta $ are real constants. Therefore, the \textit{y}-coordinate is cyclic and its corresponding momentum 
\textit{k} is conserved which then allows us to write the 4-component spinor in 3+1 dimensions as 
\begin{equation} \label{GrindEQ__3_}
\psi (t,\vec{r})=e^{i(ky-Et)} \phi (x,z).
\end{equation}
We take the following standard representation of the gamma matrices:
$$\gamma ^{0} =\left(%
\begin{array}{cc}
 1 & 0 \\
  0 & 1  \\
\end{array}%
\right), \qquad
\vec{\gamma }=\left(%
\begin{array}{cc}
  0 & \vec\si \\
  -\vec\si &  0\\
\end{array}%
\right).$$

Now after multiplying Eq. \eqref{GrindEQ__1_} by $\gamma ^{2} $, we obtain
\begin{equation} \label{GrindEQ__4_}
\left[k-ax+bz-m\gamma ^{2} -\gamma ^{0} \gamma ^{2} (E+V)-i\gamma ^{1} \gamma ^{2} \partial _{x} +i\gamma ^{2} \gamma ^{3} \partial _{z} \right]\varphi =0.
\end{equation}
Thus, the combination of matrices that appear in this equation are given by
$$\gamma ^{1} \gamma ^{2} =
-i\sigma _{3} \left(\begin{array}{cc}
 1 & 0 \\
 0 & 1  \\
\end{array} \right), \qquad \gamma ^{2} \gamma ^{3} =-i\sigma _{1} \left(\begin{array}{cc}
 1 & 0 \\
 0 & 1  \\
\end{array} \right),
\qquad \gamma ^{0} \gamma ^{2} =\si_2 \left(\begin{array}{cc}
 0 & 1 \\
 1 & 0  \\
\end{array} \right)$$
and the Dirac equation can be written as follows
\begin{equation} \label{GrindEQ__5_}
\left(\begin{array}{cc} {{\rm {\mathcal A}}} & {-{\rm {\mathcal B}}_{+} \sigma _{2} } \\ {-{\rm {\mathcal B}}_{-} \sigma _{2} } & {{\rm {\mathcal A}}} \end{array}\right)\left(\begin{array}{c} {\phi ^{+} } \\ {\phi ^{-} } \end{array}\right)=0
\end{equation}
where ${\rm {\mathcal A}}=k-ax+bz-\sigma _{3} \partial _{x} +\sigma _{1} \partial _{z} $ and ${\rm {\mathcal B}}_{\pm } =E+V\pm m$, $\phi ^{\pm } $ represent the upper and lower spinor components. Consequently, we obtain
\begin{equation} \label{GrindEQ__6_}
{\rm {\mathcal A}}\phi ^{+} -{\rm {\mathcal B}}_{+} \sigma _{2} \phi ^{-} =0
\end{equation}
\begin{equation} \label{GrindEQ__7_}
-{\rm {\mathcal B}}_{-} \sigma _{2} \phi ^{+} +{\rm {\mathcal A}}\phi ^{-} =0.
\end{equation}
Multiplying (\ref{GrindEQ__7_}) by $\sigma _{2} $, we get
\begin{equation} \label{GrindEQ__8_}
-{\rm {\mathcal B}}_{-} \phi ^{+} +{\rm {\mathcal C}}\sigma _{2} \phi ^{-} =0
\end{equation}
where ${\rm {\mathcal C}}=k-ax+bz+\sigma _{3} \partial _{x} -\sigma _{1} \partial _{z} $. Combining (\ref{GrindEQ__6_}) and \eqref{GrindEQ__8_} gives
$$\left(\begin{array}{cc} {{\rm {\mathcal A}}_{+} } & {-{\rm {\mathcal B}}_{+} } \\ {-{\rm {\mathcal B}}_{-} } & {{\rm {\mathcal A}}_{-} }
\end{array}\right)\left(\begin{array}{c} {\phi ^{+} } \\ {\sigma _{2} \phi ^{-} }
\end{array}\right)=0$$
where ${\rm {\mathcal A}}_{\pm } =k-ax+bz\mp \sigma _{3} \partial _{x} \pm \sigma _{1} \partial _{z} $. Now, let us write $\chi =\left(\begin{array}{c} {\chi ^{+} } \\ {\chi ^{-} } \end{array}\right)=\left(\begin{array}{c} {\varphi ^{+} } \\ {\sigma _{2} \varphi ^{-} } \end{array}\right)$, then this equation becomes
\begin{equation} \label{GrindEQ__9_}
\left(\begin{array}{cc} {{\rm {\mathcal A}}_{+} } & {-{\rm {\mathcal B}}_{+} } \\ {-{\rm {\mathcal B}}_{-} } & {{\rm {\mathcal A}}_{-} } \end{array}\right)\left(\begin{array}{c} {\chi ^{+} } \\ {\chi ^{-} } \end{array}\right)=0
\end{equation}
giving
\begin{equation} \label{GrindEQ__10_}
\chi ^{\pm } =\frac{1}{{\rm {\mathcal B}}_{\mp } } {\rm {\mathcal A}}_{\mp } \chi ^{\mp }.
\end{equation}
Substituting \eqref{GrindEQ__5_} into \eqref{GrindEQ__4_}, we obtain
$$\left[{\rm {\mathcal A}}_{\pm } \frac{1}{{\rm {\mathcal B}}_{\mp } } {\rm {\mathcal A}}_{\mp } -{\rm {\mathcal B}}_{\pm } \right]\chi ^{\mp } =0.$$
If the time component of the electromagnetic potential vanishes (i.e., \textit{V} = 0), then this equation becomes $\left({\rm {\mathcal A}}_{\pm } {\rm {\mathcal A}}_{\mp } -{\rm {\mathcal B}}_{\pm } {\rm {\mathcal B}}_{\mp } \right)\chi ^{\mp } =0$. More explicitly, it reads as follows
\begin{equation} \label{GrindEQ__11_}
\left[\partial _{x}^{2} +\partial _{z}^{2} -(k-ax+bz)^{2} \pm a\sigma _{3} \pm b\sigma _{1} +E^{2} -m^{2} \right]\chi ^{\pm } (x,z)=0
\end{equation}
which tell us that actually our problem is reduced to a second order differential equation. 
This two dimensional equation will be simplified and solved analytically in the following section.

%%%%%%%%%%%%%%%%%%%%%%%%%%%%%%%%%%%%%%%%%%%%%%%%%%%%%%%%%%%%%%
\section{Energy spectrum and solution space}
%%%%%%%%%%%%%%%%%%%%%%%%%%%%%%%%%%%%%%%%%%%%%%%%%%%%%%%%%%%%%%%

We look for the solution of \eqref{GrindEQ__11_} in order to explicitly determine 
the eigenvalues and corresponding eigenspinors. To do this task, 
 %To solve the last differential equation \eqref{GrindEQ__11_}, 
 we start by making the following coordinates transformation: $(x,z)\to (u,v)$
\begin{equation} \label{GrindEQ__12_}
u=bz-ax+k,\quad \quad v=az+bx+k.
\end{equation}
The new \textit{u} variable is exhibited explicitly in equation \eqref{GrindEQ__11_} while the \textit{v} variable is chosen so as to ensure that the Jacobian of the transformation \eqref{GrindEQ__12_} is constant and equal to $a^{2} +b^{2} =B^{2} $. This proves to be a simplifying hypothesis for our computations. Indeed, it results in having $\partial _{x} =$ $-a\partial _{u} +b\partial _{v} $, $\partial _{z} =b\partial _{u} +a\partial _{v} $, $\partial _{x}^{2} +\partial _{z}^{2} =B^{2} (\partial _{u}^{2} +\partial _{v}^{2} )$, and equation \eqref{GrindEQ__11_} takes the following simple form
\begin{equation} \label{GrindEQ__13_}
\left[\partial _{u}^{2} +\partial _{v}^{2} -\frac{u^{2} }{B^{2} } \pm \frac{1}{B^{2} } \left(a\sigma _{3} +b\sigma _{1} \right)+\frac{E^{2} -m^{2} }{B^{2} } \right]\chi ^{\pm } (u,v)=0.
\end{equation}

The form of the above equation suggests that it is separable in the $(u,v)$ coordinates such that we can write $\chi ^{\pm } (u,v)=e^{iR_{\pm } v} \xi ^{\pm } (u)$, where $R_{\pm } $ are two constants of length dimension and $\xi ^{\pm } =\left(\begin{array}{c} {\xi _{\uparrow }^{\pm } } \\ {\xi _{\downarrow }^{\pm } } \end{array}\right)$. Therefore, the above equation reduces to
\begin{equation} \label{GrindEQ__14_}
\left[\frac{d^{2} }{du^{2} } -\frac{u^{2} }{B^{2} } \pm \frac{1}{B} \left(\sigma _{3} \cos \theta +\sigma _{1} \sin \theta \right)+\frac{E^{2} -m^{2} }{B^{2} } -R_{\pm }^{2} \right]\xi ^{\pm } (u)=0.
\end{equation}
All terms within the bracket are diagonal except for the term containing $\sigma _{1}$, 
which results in a coupling of the two components, $\xi _{\uparrow }^{\pm } $ and $\xi _{\downarrow }^{\pm } $. 
This coupling complicates the solution of Eq. \eqref{GrindEQ__14_}. However, we rewrite it in the following alternative form
\begin{equation} \label{GrindEQ__15_}
\left[\frac{d^{2} }{du^{2} } -\frac{u^{2} }{B^{2} } \pm \frac{\sigma _{3} }{B} e^{i\theta \sigma _{2} } +\frac{E^{2} -m^{2} }{B^{2} } -R_{\pm }^{2} \right]\xi ^{\pm } (u)=0.
\end{equation}
If we multiply this equation from left by $e^{\frac{i}{2} \theta \sigma _{2} } $ and note that $e^{\frac{i}{2} \theta \sigma _{2} } \sigma _{3} =\sigma _{3} {\kern 1pt} e^{-\frac{i}{2} \theta \sigma _{2} } $, then we obtain
\begin{equation} \label{GrindEQ__16_}
\left[\frac{d^{2} }{du^{2} } -\frac{u^{2} }{B^{2} } \pm \frac{\sigma _{3} }{B} +\frac{E^{2} -m^{2} }{B^{2} } -R_{\pm }^{2} \right]\zeta ^{\pm } (u)=0
\end{equation}
where $\zeta ^{\pm } (u)=e^{\frac{i}{2} \theta \sigma _{2} } \xi ^{\pm } (u)$. Since $\sigma _{3} $ is diagonal then the two components, $\xi _{\uparrow }^{\pm } $ and $\xi _{\downarrow }^{\pm } $, are completely decoupled. Clearly, Eq. \eqref{GrindEQ__16_} corresponds to an oscillator problem in the \textit{u} variable and hence we can use the related machinery to find the required solutions. Solving this equation gives the two-component spinor $\chi ^{\mp } $, which when substituted in Eq. \eqref{GrindEQ__10_} gives the other two-component spinor $\chi ^{\pm } $ that belong to the $\pm $ energy spaces. That is, we obtain the complete solution space by studying separately the two energy subspaces: one corresponds to the top signs in all equations above and the other corresponds to the bottom signs. The flowchart diagram below summarizes the scheme:
%\begin{center}
%\includegraphics*[width=3.94in, height=1.59in, keepaspectratio=false]{fig1.ps}
%\end{center}

$${{\bf Eq.(16)} \longrightarrow \zeta^{\pm} \longrightarrow\chi^{\pm} = e^{-\frac{1}{2} \te\si_2} e^{iR_{\pm}v} \zeta^{\pm}\
{}^{\nearrow\ {}^{ \chi^+\ \longrightarrow \ \mbox{\bf Eq. (10)}\ \longrightarrow\ \chi^- \ \longrightarrow\
\binom{\phi^+}{\phi^-}=\binom{\chi^+} {\si_2\chi^-}  }}
_{\searrow\ {}_ { \chi^-\ \longrightarrow \ \mbox{\bf Eq. (10)}\ \longrightarrow\ \chi^+ \ \longrightarrow\
\binom{\phi^+}{\phi^-}=\binom{\chi^+} {\si_2\chi^-} }}} $$

The solution of Eq. \eqref{GrindEQ__16_} is then given by
\begin{equation} \label{GrindEQ__17_}
\zeta _{n}^{+} (u)=e^{-u^{2} /2B} \left(\begin{array}{c} {C_{n}^{+} \hat{H}_{n} (u/\sqrt{B} )} \\ {D_{n-1}^{+} \hat{H}_{n-1} (u/\sqrt{B} )} \end{array}\right)
\end{equation}
where $C_{n}^{\pm }$ and $D_{n}^{\pm }$ are normalization constants and $\hat{H}_{n} (x)$ are the renormalized Hermite polynomials\footnote{These are the Hermite polynomials $H_{n} (x)$,
divided by  $\sqrt{\pi ^{1/2} 2^{n} n!}$  so that  $\int _{-\infty }^{+\infty } e^{-x^{2} } \hat{H}_{n} (x)\hat{H}_{m} (x) dx=\delta _{nm}$.}. The zero-energy state reads
\begin{equation} \label{GrindEQ__18_}
\zeta _{0}^{+} (u)=\pi ^{-\frac{1}{4} } C_{0}^{+} e^{-u^{2} /2B} \left(\begin{array}{c} {1} \\ {0} \end{array}\right).
\end{equation}
The corresponding energy eigenvalues are given by
\begin{equation} \label{GrindEQ__19_}
E_{n}^{+} =\pm \sqrt{m^{2} +(BR_{+} )^{2} +2nB} .
\end{equation}
On the other hand,
\begin{equation} \label{GrindEQ__20_}
\zeta _{n}^{-} (u)=e^{-u^{2} /2B} \left(\begin{array}{c} {D_{n-1}^{-} \hat{H}_{n-1} (u/\sqrt{B} )} \\ {C_{n}^{-} \hat{H}_{n} (u/\sqrt{B} )} \end{array}\right)
\end{equation}
and the corresponding vacuum reads
\begin{equation} \label{GrindEQ__21_}
\zeta _{0}^{-} (u)=\pi ^{-\frac{1}{4} } C_{0}^{-} e^{-u^{2} /2B} \left(\begin{array}{c} {0} \\ {1} \end{array}\right)
\end{equation}
where the associated eigenvalues read
\begin{equation} \label{GrindEQ__22_}
E_{n}^{-} =\pm \sqrt{m^{2} +(BR_{-} )^{2} +2nB} .
\end{equation}
Therefore, we have 
\begin{equation} \label{GrindEQ__23_}
\chi _{n}^{\pm } (u,v)=e^{-\frac{i}{2} \theta \sigma _{2} } e^{iR_{\pm } v} \zeta _{n}^{\pm } (u)
\end{equation}
and, finally, Eq. \eqref{GrindEQ__10_} gives
\begin{equation} \label{GrindEQ__24_}
\chi _{n}^{\mp } (u,v)=\frac{e^{-i\frac{\theta }{2} \sigma _{2} } }{E\pm m} {\kern 1pt} {\kern 1pt} e^{iR_{\pm } v} \left[u\pm B\left(\sigma _{3} \frac{d}{du} +iR_{\pm } \sigma _{1} \right)\right]\zeta _{n}^{\pm } (u)
\end{equation}
where the upper signs in Eqs. \eqref{GrindEQ__23_} and \eqref{GrindEQ__24_} correspond to the positive energy solutions whose energy spectrum is given by Eq. \eqref{GrindEQ__19_} with the positive sign. The negative energy solutions correspond to the lower signs in (\ref{GrindEQ__22_}-\ref{GrindEQ__24_}).

Using our results \eqref{GrindEQ__17_} and \eqref{GrindEQ__20_} in Eq. \eqref{GrindEQ__24_} and utilizing the differential property and recursion relation of the Hermite polynomials we obtain
\begin{eqnarray} 
\chi _{n}^{-} (u,v) &=&\sqrt{B} \frac{e^{-u^{2} 2B} }{E+m} {\kern 1pt} {\kern 1pt} e^{iR_{+} v} e^{-i\frac{\theta }{2} \sigma _{2} } \left(\begin{array}{c} {\left[2nC_{n}^{+} +i\sqrt{B} R_{+} D_{n-1}^{+} \right]\hat{H}_{n-1} (u/\sqrt{B} )} \\ {\left[D_{n-1}^{+} +i\sqrt{B} R_{+} C_{n}^{+} \right]\hat{H}_{n} (u/\sqrt{B} )} \end{array}\right) \label{GrindEQ__25_}\\
\chi _{0}^{-} (u,v) &=& i\pi ^{-\frac{1}{4} } \frac{BR_{+} C_{0}^{+} }{E+m} e^{iR_{+} v} e^{-u^{2} /2B} e^{-i\frac{\theta }{2} \sigma _{2} } \left(\begin{array}{c} {0} \\ {1} \end{array}\right)\label{GrindEQ__26_}\\
\chi _{n}^{+} (u,v) &=& \sqrt{B} \frac{e^{-u^{2} /2B} }{E-m} {\kern 1pt} {\kern 1pt} e^{iR_{-} v} e^{-i\frac{\theta }{2} \sigma _{2} } \left(\begin{array}{c} {\left[D_{n-1}^{-} -i\sqrt{B} R_{-} C_{n}^{-} \right]\hat{H}_{n} (u/\sqrt{B} )} \\ {\left[2nC_{n}^{-} -i\sqrt{B} R_{-} D_{n-1}^{-} \right]\hat{H}_{n-1} (u/\sqrt{B} )} \end{array}\right)\label{GrindEQ__27_}\\
\chi _{0}^{+} (u,v) &=&i\pi ^{-\frac{1}{4} } \frac{BR_{-} C_{0}^{-} }{E-m} e^{iR_{-} v} e^{-u^{2} /2B} e^{-i\frac{\theta }{2} \sigma _{2} } \left(\begin{array}{c} {-1} \\ {0} \end{array}\right).\label{GrindEQ__28_}
\end{eqnarray}
Due to the fact that
\begin{equation} \label{GrindEQ__29_}
\left(\begin{array}{c} {dx} \\ {dz} \end{array}\right)=\frac{1}{B^{2} } \left(\begin{array}{cc} {-a} & {b} \\ {b} & {a} \end{array}\right)\left(\begin{array}{c} {du} \\ {dv} \end{array}\right)=\frac{1}{B} \left(\begin{array}{cc} {-\cos \theta } & {\sin \theta } \\ {\sin \theta } & {\cos \theta } \end{array}\right)\left(\begin{array}{c} {du} \\ {dv} \end{array}\right)
\end{equation}
then, the Jacobian of the transformation matrix is $1/B^{2} $ and $dxdz=B^{-2} dudv$. Hence, the probability density (time component of the current density) becomes
\begin{equation} \label{GrindEQ__30_}
\begin{array}{rcl} 
{\left|\psi \right|^{2} } {=}  {\int \bar{\psi }\gamma ^{0} \psi\ dxdz =\int \psi ^{\dag } \psi\ dxdz =\int \left(|\phi ^{+} |^{2} +|\phi ^{-} |^{2} \right)\ dxdz }. 
\end{array}
\end{equation}
With the help of the orthogonality property of the Hermite polynomials, we can show that the energy eigenfunctions, $\left\{\psi _{n} \right\}_{n=0}^{\infty } $, are orthonormal (i.e., $\left\langle \psi _{n} \left|\right. \psi _{m} \right\rangle =\delta _{nm} $) and the normalization $\left|\psi _{n} \right|^{2} =1$ gives
\begin{equation} \label{GrindEQ__31_}
(C_{n}^{\pm } )^{2} +(D_{n-1}^{\pm } )^{2} =\lambda B\frac{E_{n}^{\pm } \pm m}{2E_{n}^{\pm } } 
\end{equation}
where the real dimensionless parameter $\lambda =L\sqrt{B}$ and \textit{L} is the characteristic size of the physical system, which is defined by $L^{-1} =\int  dv$
while the magnetic length is $\ell _{B} \approx 1/\sqrt{B}$ so that $\lambda $ is just the ratio of the system size to the magnetic length.

%%%%%%%%%%%%%%%%%%%%%%%%%%%%%%%%%%%%%%
\section{Discussions}
%%%%%%%%%%%%%%%%%%%%%%%%%%%%%%%%%%%%%%%%

Now let us make some comments about the results obtained so far. In the normalization condition, Eq. \eqref{GrindEQ__31_}, we notice that
  $\left(E_{n}^{\pm } \pm m\right)E_{n}^{\pm } >0$ so that the right hand side of this equation is always positive as required. Moreover, 
  we end up with only two independent normalization constants, say $C_{n}^{\pm } $. The energy dispersion relation
\begin{equation} \label{GrindEQ__32_}
E_{n}^{\pm } =\pm \sqrt{m^{2} +(BR_{\pm } )^{2} +2nB} 
\end{equation}
reduces to the massless Dirac particle in a perpendicular magnetic field if $R_{\pm } $ = 0 and $m=0$ as observed in spectroscopic experiments on graphene 
layers \cite{5} and its unusual manifestation in the anomalous quantum Hall effect \cite{222,2222,6}, as observed experimentally. We also notice that 
the presence of a transverse component of the magnetic field gives rise to an effective mass term in \eqref{GrindEQ__32_} which is proportional to the magnitude of the magnetic field.

The above analysis suggests that we should inspect two limiting cases and see what one can gain from the form of the energy spectrum. First, 
for a strong magnetic field situation, we can show that the energy solutions can be reduced to the form
\begin{equation} \label{GrindEQ__33_}
E_{n}^{\pm } \approx \pm \frac{1}{R_{\pm } } \left(n+BR_{\pm }^{2} \right).
\end{equation}
 These can be regarded as the Landau energy levels for a particle moving perpendicular to the uniform magnetic field. More precisely, the problem is reduced to a harmonic oscillators where the parameter ${1\mathord{\left/ {\vphantom {1 R_{\pm } }} \right. \kern-\nulldelimiterspace} R_{\pm } } $ can be interpreted as a frequency of oscillation and the terms $BR_{\pm } $ as shift in the energy levels. Clearly, the energy spacing is now measured in terms of ${1\mathord{\left/ {\vphantom {1 R_{\pm } }} \right. \kern-\nulldelimiterspace} R_{\pm } } $, which is nothing but a gap between two successive levels.

Second, let us consider the weak magnetic field limit. After some calculation, we end up with the energy spectrum %reduces to
\begin{equation} \label{GrindEQ__34_}
E_{n}^{\pm } \approx \pm \sqrt{m^{2} +2nB} .
\end{equation}
What is remarkable in this limit is that we lost the dependence on $R_{\pm } $ and the system behaves like that of massive Dirac particles confined 
in a plane perpendicular to the magnetic field. These limiting results show clearly that our system is rich and can be used to extract more interesting 
information, which can be applied to both massive and massless Dirac particles.

%%%%%%%%%%%%%%%%%%%%%%%%%%%%%%%%%%%%%%
\section{Conclusion}
%%%%%%%%%%%%%%%%%%%%%%%%%%%%%%%%%%%%%%%%

 In this paper, we have obtained analytically the bound states solution of the Dirac equation in 3+1 dimensions in the presence of a 
 uniform magnetic field background using a particular gauge, as given in Eq. \eqref{GrindEQ__2_}. From the derived dispersion relation 
 we see the emergence of Landau energy levels reflecting the quantized nature of the particle motion. However, the obtained closed form 
 solutions dependent on the choice of gauge. In general, the quantities calculated in different gauges differ by the choice of coordinate 
 axes. However, physical quantities (e.g., scattering cross-section, energy spectrum, etc.) are gauge independent as required on physical 
 grounds. Thus in our present problem the fact that the spinor solutions are gauge dependent is not alarming since all unphysical appearances
 could be gauged away.

 Before closing, we comment on some potential applications of our results. One is to study the tunneling effect through the evaluation of the 
 reflection and transmission coefficients of a propagating fermions along the vector potential (the \textit{y}-axis). The fact that our computations 
 spans both massive and massless cases can be used to study Klein tunneling in these systems and use the tilt angle as a parameter, which will enable
 us to move from one regime to the other. More importantly, the extension of our findings to describe some features in graphene systems will be desirable, 
 in particular concerning the anomalous quantum Hall conductivity and related matter.

%%%%%%%%%%%%%%%%%%%%%%%%%%%%%%%%%%%%%%%%%%
\section*{Acknowledgments}
%%%%%%%%%%%%%%%%%%%%%%%%%%%%%%%%%%%%%%%%%%%%%%%

The generous support provided by the Saudi Center for Theoretical Physics (SCTP) is highly appreciated by all Authors.
We also acknowledge the support provided by King Fahd University of Petroleum $\& $ Minerals under project RG1108-1-2.

%%%%%%%%%%%%%%%%%%%%%%%%%%%%%%%%%%%%%%
\section*{Appendix}
%%%%%%%%%%%%%%%%%%%%%%%%%%%%%%%%%%%%%%%%

In this appendix we use a different gauge that describes the same uniform magnetic field. In this new gauge the space component of the static electromagnetic potential is given by
\beq
\vec{A}(\vec{r})=(0,a\, x,by) \tag{A.1}                                                                               %(A.1)
\eeq
where \textit{a} and \textit{b} are as defined below Eq. \eqref{GrindEQ__2_}. Therefore, the \textit{z} coordinate is cyclic and we can write the 4-component spinor in 3+1 dimensions as $\psi (t,\vec{r})=e^{i(kz-Et)} \phi (x,y)$. Multiplying the Dirac equation \eqref{GrindEQ__1_} from left by $\gamma ^{3} $, we obtain
\beq
\left[k-by-m\gamma ^{3} -\gamma ^{0} \gamma ^{3} (E+V)+i\gamma ^{3} \gamma ^{1} \partial _{x} -\gamma ^{2} \gamma ^{3} (i\partial _{y} +ax)\right]\phi =0.           \tag{A.2}
\eeq
 
 Taking the standard representation of the gamma matrices as shown below Eq. \eqref{GrindEQ__3_}, we obtain the matrices
$$\gamma ^{3} \gamma ^{1} =-i\sigma _{2} \left(\bra{cc} 1& 0 \\ 0 & 1 \era\right), \qquad
\gamma ^{2} \gamma ^{3} =-i\sigma _{1} \left(\bra{cc} 1& 0 \\ 0 & 1 \era\right),
\qquad \gamma ^{0} \gamma ^{3} = \sigma _{3}\left(\bra{cc}0 & 1  \\ 1 &0 \era \right).$$
Thus, the Dirac equation (A.2) can be written as follows
\beq
 \left(\begin{array}{cc} {\cal A} & {-{\cal B}_{+} \sigma _{3} } \\ {-{\cal B}_{-} \sigma _{3} } & {\cal A} \end{array}\right)\left(\begin{array}{c} {\phi ^{+} } \\ {\phi ^{-} } \end{array}\right)=0
 \tag{A.3}                                                                  %(A.4)
\eeq
where ${\rm {\mathcal A}}=k-by+\sigma _{2} \partial _{x} -\sigma _{1} (\partial _{y} -iax)$ and ${\rm {\mathcal B}}_{\pm } =E+V\pm m$. Consequently, we obtain
\beq
{\rm {\mathcal A}}\phi ^{+} -{\rm {\mathcal B}}_{+} \sigma _{3} \phi ^{-} =0                                                                                             \tag{A.4}
\eeq
\beq
-{\rm {\mathcal B}}_{-} \sigma _{3} \phi ^{+} +{\rm {\mathcal A}}\phi ^{-} =0.                                                                                            \tag{A.5}
\eeq
Multiplying \eqref{GrindEQ__2_} by $\sigma _{3} $ we obtain
\beq
-{\rm {\mathcal B}}_{-} \phi ^{+} +{\rm {\mathcal C}}\sigma _{3} \phi ^{-} =0                                                                                           \tag{A.6}
\eeq
where ${\rm {\mathcal C}}=k-by-\sigma _{2} \partial _{x} +\sigma _{1} (\partial _{y} -iax)$. Combining (A.4) and (A.5) gives
\beq
\left(\begin{array}{cc} {{\rm {\mathcal A}}_{+} } & {-{\rm {\mathcal B}}_{+} } \\ {-{\rm {\mathcal B}}_{-} } & {{\rm {\mathcal A}}_{-} } \end{array}\right)\left(\begin{array}{c} {\phi ^{+} } \\ {\sigma _{3} \phi ^{-} } \end{array}\right)=0                                                                                   \tag{A.7}
\eeq
where ${\rm {\mathcal A}}_{\pm } =k-by\pm \sigma _{2} \partial _{x} \mp \sigma _{1} (\partial _{y} -iax)$. 
Now, let us write $\chi =\left(\bra{c} \chi^{+} \\ \chi^{-} \era\right)=\left(\bra{c} \phi^+ \\ \sigma _{3} \phi^{-} \era \right)$, then this equation becomes
\beq
\left(\begin{array}{cc} {{\rm {\mathcal A}}_{+} } & {-{\rm {\mathcal B}}_{+} } \\ {-{\rm {\mathcal B}}_{-} } & {{\rm {\mathcal A}}_{-} } 
\end{array}\right)\left(\begin{array}{c} {\chi ^{+} } \\ {\chi ^{-} } \end{array}\right)=0                                                                                     \tag{A.8}
\eeq
giving the relation
\beq
\chi ^{\pm } =\frac{1}{{\rm {\mathcal B}}_{\mp } } {\rm {\mathcal A}}_{\mp } \chi ^{\mp }.                                                                                        \tag{A.9}
\eeq
Substituting (A.9) into (A.8), we obtain
\beq
\left[{\rm {\mathcal A}}_{\pm } \frac{1}{{\rm {\mathcal B}}_{\mp } } {\rm {\mathcal A}}_{\mp } -{\rm {\mathcal B}}_{\pm } \right]\chi ^{\mp } =0.                                                                                  \tag{A.10}
\eeq

If the time component of the electromagnetic potential vanishes (i.e., \textit{V} = 0), then this equation becomes 
$\left({\rm {\mathcal A}}_{\pm } {\rm {\mathcal A}}_{\mp } -{\rm {\mathcal B}}_{\pm } {\rm {\mathcal B}}_{\mp } \right)\chi ^{\mp } =0$. In details, it reads
\beq
\left[\partial _{x}^{2} +\partial _{y}^{2} -2iax\partial _{y} -a^{2} x^{2} -(by-k)^{2} +a\sigma _{3} \pm b\sigma _{1} +E^{2} -m^{2} \right]\chi ^{\pm } (x,y)=0.
 \tag{A.11}
\eeq
This equation clearly shows that $x$ and $y$ degrees of freedom are coupled and the two components of $\chi ^{\pm } (x,y)$ are coupled through the magnetic field 
term $b$. Hence, this coupling is suppressed for perpendicular magnetic fields ($\theta $ = 0). The above details show that even though the physics is 
invariant under gauge transformation the mathematics can become really messy in the wrong gauge so to speak.

At this stage let us introduce a connection between both gauges. This can be worked out by using the standard approach to find the suitable gauge transformation. 
Indeed, we transform the spinors and potential as follows
\beq
\chi (\vec{r})\to e^{if(\vec{r})} \chi (\vec{r}),\quad \quad A_{\mu } (\vec{r})\to A_{\mu } (\vec{r})+\partial _{\mu } f(\vec{r})   \tag{A.12}
\eeq
where $f(\vec{r})$ is an arbitrary scalar function. According to the two gauges considered in our problem, we can show that such a function takes the following form
\beq
f(\vec{r})=f(y,z)=b\, y\, z.       \tag{A.13}
\eeq

\end{document}